\documentclass[
    a4paper,
    amsmath,
    amssymb,
    aps
]{revtex4}

\usepackage[T1]{fontenc}
\usepackage[utf8]{inputenc}

\usepackage{amsmath}
\usepackage{amssymb}
\usepackage{bm}
\usepackage{relsize}

\usepackage{graphicx}
\usepackage{xcolor}
\usepackage{tikz}
\usetikzlibrary{trees}

\usepackage{dcolumn}

\usepackage[margin=1in]{geometry}
\usepackage[section]{placeins}

\let\Oldsection\section
\renewcommand{\section}{\FloatBarrier\Oldsection}

\let\Oldsubsection\subsection
\renewcommand{\subsection}{\FloatBarrier\Oldsubsection}

\let\Oldsubsubsection\subsubsection
\renewcommand{\subsubsection}{\FloatBarrier\Oldsubsubsection}

\usepackage{setspace}

\usepackage{graphicx}
\graphicspath{{Figures/}}
\begin{document}
\setstretch{1.5}
\preprint{APS/123-QED}
\title{Towards a quantum decision tree in a laser pumped four-level system}

\author{Dawit Hiluf Hailu}
\email{dhailu@bowiestate.edu}
\affiliation{Department of Natural Sciences, Bowie State University, 14000 Jericho Park Road, Bowie, 20715, USA}

\date{\today}
\begin{abstract}
In this study, we examine an innovative framework towards implementing quantum decision trees utilizing a laser-driven four-level system. We discuss a diamond-shaped atomic configuration, in which we apply Lie-algebraic formalisms to analyze the dynamics of the system. The system is perturbed by a Stokes pulse, represented as $\beta_j(t)$ (for $j=1,2$), which interacts with the atomic states $|0\rangle, |3\rangle$ and $|1\rangle, |2\rangle$. In addition, a pump laser, denoted as $\alpha_j(t)$, couples the states $|0\rangle, |1\rangle$ and $|2\rangle, |3\rangle$.
By employing pulse profiles that possess identical temporal behavior but differ in amplitude, one can effectively redistribute the population from the initial ground state to the other energy levels. This technique facilitates the mimicry of a quantum decision tree. We highlight that the proposed methodology is scalable to N-level systems, enhancing its adaptability and potential utility in quantum computing and various decision-making applications.

We introduce a novel framework for implementing quantum decision trees using a four-level laser-driven atomic system. Employing a diamond-shaped energy configuration, we analyze system dynamics through Lie-algebraic methods. The system's interactions are mediated by Stokes pulses, \(\beta_j(t)\) (for \(j=1,2\)), coupling states \(|0\rangle \leftrightarrow |3\rangle\) and \(|1\rangle \leftrightarrow |2\rangle\), alongside pump pulses, \(\alpha_j(t)\), facilitating transitions between \(|0\rangle \leftrightarrow |1\rangle\) and \(|2\rangle \leftrightarrow |3\rangle\). Using pulse profiles with identical temporal structures but varying amplitudes, we achieve controlled population redistribution among energy levels, effectively simulating a quantum decision tree. This methodology is scalable to systems of \(N\) levels, offering potential applications in quantum computing and decision-making processes.

\end{abstract}

\maketitle


\section{Introduction}

In computational complexity theory, decision trees are a fundamental model of computation, providing a framework for understanding how algorithms operate through a sequence of adaptive queries. Each query is influenced by the results of the previous tests, which direct the algorithm along different branches of the tree until a final decision or result is reached. This model is crucial for analyzing the complexity of decision-making processes in various computational problems\cite{BuhrmandeWolf2002}.

A binary decision tree can be implemented in a molecular system, as demonstrated by 2D spectroscopy techniques \cite{Fresch:2013aa,FreshMDT}. The electrical counterpart, a multivariate decision tree, has been realized through a dopant molecule embedded in silicon \cite{FreschQDT}. A decision tree can be viewed as a visual representation of the truth table of a logic function, offering a straightforward way to illustrate the relationships between inputs and outputs. For example, consider the OR gate: a simple two-input logic gate that produces a true output whenever at least one of its inputs is true.

The truth table for the OR gate outlines all possible combinations of its two inputs and their respective outputs. With each input 0 or 1, there are \(2^2 = 4\) combinations to consider \cite{kohavi2010switching}. This truth table can be mapped onto a binary decision tree that consists of two levels of decision points plus a third level containing the terminal nodes. Each nonterminal node in the tree evaluates an input variable as true (1) or false (0).

Starting at the root, the decision tree first assesses the value of the first input. If this input is true (1), the path moves to the left child node; if false (0), it moves to the right. At the next level, the second input is evaluated similarly, guiding the path toward a terminal node (or leaf). These terminal nodes represent the output of the OR gate for each input combination. Given that the OR gate has two inputs, the decision tree spans two levels of decisions, resulting in four possible arrangements that lead to the four terminal nodes.

Extending this concept into the quantum realm, we propose the emulation of a quantum-mechanical decision tree. Here, an initial quantum state, positioned at the root, evolves through the tree structure by exploring multiple branches simultaneously due to quantum superposition and interference principles. This quantum approach can potentially enable faster evaluation of decision paths based on input variables, leveraging the inherent parallelism of quantum states to make efficient computations.

Classical decision trees, widely used in machine learning and artificial intelligence, operate on binary structures that limit their efficiency in handling complex decision-making tasks. Although quantum machine learning has been proposed as a novel approach, quantum decision trees can be considered a subset of this field. However, most of the research has focused on qubit-based platforms such as superconducting circuits and trapped ions \cite{QMLBiamonte,QMLSchuld03042015,QMLTMonz2016}. However, their implementation in atomic systems, particularly those based on four-level atomic structures, remains largely unexplored.  

In this work, we present a novel proposal toward the implementation of a quantum decision tree using a four-level laser-pumped atomic system, effectively replacing the classical binary decision tree framework. Using controlled atomic transitions and quantum coherence, our approach processes decision paths beyond binary constraints, enabling massively parallel computations. Unlike conventional methods, our framework extends the Bloch sphere representation utilizing an expanded generator set, which decomposes the system evolution on an orthonormal basis. Additionally, we suggest the use of rare-earth-ion-doped crystals, which are recognized for their extended coherence times, to improve stability and reduce decoherence-related errors. This approach positions our method as a strong candidate for practical quantum computing applications.  


\begin{figure}[!htbp]
\centering
\begin{minipage}{0.45\textwidth}
\centering    
\begin{tikzpicture}
  [scale=1.0, 
   level distance=20mm, 
   every node/.style={circle, minimum size=7mm, inner sep=1pt, font=\small}, 
   level 1/.style={sibling distance=40mm, nodes={fill=blue!45}}, 
   level 2/.style={sibling distance=10mm, nodes={fill=green!30}}, 
   edge from parent/.style={draw, thick}]
  
  \node[fill=red!60] (root) {S} 
    child {node (A0) {S}
      child {node (A0B0) {0}} 
      child {node (A0B1) {1}} 
    }
    child {node (A1) {S}
      child {node (A1B0) {1}} 
      child {node (A1B1) {1}}  
    };

  \path[draw=blue] (root) -- (A0) node[midway, left] {0}; 
  \path[draw=red] (root) -- (A1) node[midway, right] {1}; 
  \path[draw=blue] (A0) -- (A0B0) node[midway, left] {0}; 
  \path[draw=red] (A0) -- (A0B1) node[midway, right] {1}; 
  \path[draw=blue] (A1) -- (A1B0) node[midway, left] {0}; 
  \path[draw=red] (A1) -- (A1B1) node[midway, right] {1}; 

\end{tikzpicture}
\caption{Decision tree representation of the OR logic gate. In this tree, the path taken through the nodes represents the decision-making process, where each path corresponds to evaluating the OR gate for a specific input combination. The paths corresponding to input 0 are shown in blue, while those for input 1 are shown in red.}
\end{minipage}%
\hfill
\begin{minipage}{0.5\textwidth}
\centering
\begin{tabular}{|c|c|c|}
\hline
Input 1 & Input 2 & Output \\
\hline
0 & 0 & 0 \\
0 & 1 & 1 \\
1 & 0 & 1 \\
1 & 1 & 1 \\
\hline
\end{tabular}
\caption{Truth table for the OR logic gate.}
\end{minipage}
\end{figure}
\FloatBarrier
The outline of the paper is structured as follows: Section (\ref{sec:SysHamilt}) introduces the four-level diamond-shaped atomic system under consideration, detailing its specific configuration and the interactions involved. The Hamiltonian governing the system's dynamics is also presented, setting the foundation for subsequent analysis. Next, Section (\ref{sec:SUND}) presents the dynamics of the $\mathrm{SU(4)}$ group using the Alhassid-Levine (AL) formalism. Section (\ref{sec:EQOM}) then derives the equations of motion for the four-level system using the $\mathrm{SU(4)}$ group. This derivation involves embedding the atomic variables within the Lie algebra framework, resulting in a set of coupled differential equations that describe the system's temporal evolution. Section (\ref{sec:noise}) provides a discussion on the incorporation of noise and its impact on the dynamics. Section (\ref{sec:decsiontree}) presents a path toward implementing a quantum decision tree in an optically pumped four-level system. Finally, Section (\ref{sec:conclusion}) concludes the article with a summary of the key findings and their implications. 
\section{Four Level System}
\label{sec:SysHamilt}
\begin{figure}[!htb]
\begin{center}
\includegraphics[width=2.5 in]{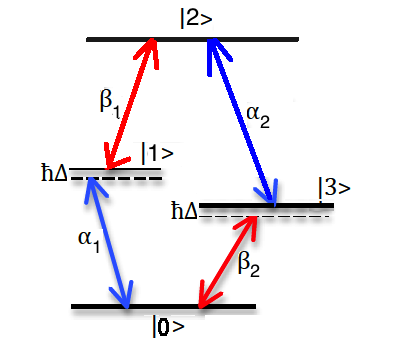}
\caption{Four-level diamond-shaped, with coupling lasers. Blue pump laser, and red Stokes laser.}
\label{4level}
\end{center}
\end{figure}

The system we consider is shown in Fig.~(\ref{4level}). It is a four-level system in a diamond shape, with states $|0\rangle$, $|1\rangle$, $|2\rangle$, and $|3\rangle$. The states $|0\rangle,|3\rangle$ and $|1\rangle,|2\rangle$ coupled by the Stokes laser, $\beta_2\left(t\right), \beta_1\left(t\right)$ respectively; and the states $|0\rangle,|1\rangle$ and $|2\rangle,|3\rangle$ coupled by the pump laser $\alpha_1\left(t\right),\alpha_2\left(t\right)$, respectively. The detunings are given by $\Delta_{\alpha}=\omega_{\alpha}-\omega_{01}$ and $\Delta_{\beta}=\omega_{\beta}-\omega_{30}$ with $\omega_{nm}=\omega_{n}-\omega_{m}$. There is no coupling between the states $|1\rangle,|3\rangle,$ and there is no direct transition between states $|0\rangle,|2\rangle$ \cite{Shore:1992aa,Shore:2006aa,SHORE:1991aa,Shore:2008aa,Garcia-Fernandez:2006aa}. Stokes pulses play a crucial role in the dynamics by facilitating transitions between energy levels in the four-level system. These pulses help control the population distribution among the quantum states and enable the creation of coherent superposition states essential for quantum decision-making. Throughout this paper, we consider the case where $\Delta_{\alpha}=\Delta_{\beta}=\Delta$
\graphicspath{{Figures//}}


Quantum systems with discrete states, such as the four-level system examined here, are known to undergo state transitions when subjected to external perturbations \cite{levine2011quantum,shore98,fano1957description}. The unperturbed Hamiltonian, $\hat{H}_0$, defines the eigenstates of the system, labeled $|0\rangle$, $|1\rangle$, $|2\rangle$, and $|3\rangle$. Interaction with a laser field introduces an additional term $\hat{H}_I$, which is related to the dynamics of the system. As a result, the total Hamiltonian can be written as the sum $\hat{H} = \hat{H}_0 + \hat{H}_I$. By adopting the interaction picture and applying the rotating wave approximation (RWA), this expression can be further refined \cite{shore2011manipulating,Bergmann2001,Berman1998}:
 \begin{equation}
 \begin{aligned}
\hat H=
 \frac{\hbar}{2}\begin{pmatrix}
  0 & \alpha_1\left(t\right)  & 0 & \beta_2\left(t\right) \\
 \alpha_1\left(t\right) & 2\Delta & \beta_1\left(t\right) & 0 \\
  0 & \beta_1\left(t\right) & 0 & \alpha_2\left(t\right)\\
  \beta_2\left(t\right) & 0 & \alpha_2\left(t\right) & 2\Delta
 \end{pmatrix}
 \end{aligned}
\end{equation}
The Hamiltonian provided describes a four-level quantum system interacting with a time-dependent external field. The diagonal elements represent the energy levels of the system's states in the absence of interactions, while the off-diagonal elements capture the time-dependent couplings between different states\cite{HailuFSM}. These couplings illustrate how an external laser field drives transitions between the states, with the strengths of the interactions changing over time. Typically employed in the interaction picture with the rotating wave approximation (RWA), this form of the Hamiltonian facilitates analysis of the system's evolution as it responds to external perturbations. In the following section, we derive the equations of motion for this four-level system using the \(\mathrm{SU(4)}\) group, allowing for a detailed treatment of the atomic variables and their dynamics.
\section{$\mathrm{SU(4)}$ Dynamics}
\label{sec:SUND}
The density matrix formalism serves as a fundamental tool for analyzing quantum dynamics in complex systems \cite{fano1957description, Eberly1975, blum2012density, breuer2007theory}. This method calculates the time evolution of a quantum system by averaging over expectation values, thus providing insight into observable macroscopic properties such as populations and coherences. The physical interpretation of each element of the density matrix is strongly influenced by the chosen representation. In this work, we adopt a representation where the unperturbed Hamiltonian \(H_0\) is diagonal. In this representation, the elements of the density matrix are expressed as \(\rho_{jk} = \langle \psi_j | \hat{\rho} | \psi_k \rangle\), where \(\hat{\rho}\) denotes the density operator.

In this representation, the diagonal elements \(\rho_{jj}\) correspond to the ensemble-averaged occupation probabilities of the quantum state \(|\psi_j\rangle\), while the off-diagonal elements \(\rho_{jk}\) reveal the averaged relative phases between states, commonly referred to as coherences \cite{CohenOBE, mukamel1999principles}. The Hermitian nature of the density matrix \(\rho\) means that the dynamics of the populations \(\rho_{jj}\) is inherently linked to the dynamics of the coherences \(\rho_{jk}\). Therefore, the density matrix must be considered as an integrated entity to fully understand the evolution of the system \cite{mukamel1999principles}. For the four-level system under study, the \(4 \times 4\) Hermitian density matrix \(\rho\) involves sixteen real parameters, of which only 15 are independent due to the normalization condition that ensures the conservation of the total probability \cite{MichaelChuang2010}.

The observable vector can be conceptualized as an advanced interpretation of the Bloch vector \cite{Fresch:2013aa, Hiluf_2018, hioe1985two, dattoli1991matrix, Aravind:86}, enhancing our ability to visualize quantum states. This approach enables the representation of a quantum state within a closed 15-dimensional space, similar to how the Bloch sphere encapsulates quantum states in three dimensions. This conceptual framework will be pivotal in guiding our subsequent analysis \cite{HIOE:1981aa, hioe1982nonlinear, hioe1983dynamic}.

Furthermore, the Optical Bloch Equation \cite{feynman1957geometrical, Blochpackard1946nuclear,dattoli1988evolution}, which is initially formulated for two-level systems, can be adapted to investigate the dynamics of $N$-level systems. In this context, these systems are represented as vectors within a $(N^2 - 1)$-dimensional space, utilizing the $\mathrm{SU(N)}$ symmetry to represent atomic variables as vectors with $N^2 - 1$ components. The work of Alhassid-Levine (AL) \cite{alhassid1977entropy, AlhassidLevine} and Hioe-Eberly (HE) \cite{HIOE:1981aa, hioe1982nonlinear, hioe1983dynamic, DattoliPRA, Aravind:86} has further extended the use of Lie algebra to explore the dynamics of $N$-level systems.

To commence our discussion, we will first outline the derivations pertinent to this methodology. The generators, denoted as $\hat{G}_{\alpha}$, are formulated as linear combinations of the operators $|j\rangle \langle k|$, where $j, k$ range from 1 to 4 \cite{Fresch:2013aa,DnstyQMDynamicsKsenia}. Although there are originally sixteen generators, normalization reduces this to a total of fifteen; their explicit definitions can be found in Eq. \eqref{gene}. 
In addition, the generators \(\hat{G}_{\alpha}\) possess the following properties: 
\begin{subequations}
\begin{align}
Tr\left(\hat G_{\alpha}\hat G_{\beta}\right)= &2\delta_{\alpha\beta}\label{trace}\\
\left[\hat G_{\alpha} ,\hat G_{\beta}\right]=&2i f_{\alpha\beta\gamma} \hat G_{\gamma}\label{comuG}
\end{align}
\end{subequations}
Where repeated indices are summed from $1$ to $15$, and $f_{\alpha\beta\gamma}$ denotes the completely antisymmetric structure constants of the SU(4) group, which are detailed in Table \ref{su4c} in Appendix \ref{su4}. These relations establish that the generators form an orthonormal set under the trace inner product and obey the Lie algebra commutation relations. 
We will utilize these generators to appropriately expand our Hamiltonian and density matrix.
\begin{subequations}
\begin{align}
\hat\rho\left(t\right)=&\frac{\hat I}{4}+\frac{1}{2}\sum_{\alpha=1}^{15} \langle G_{\alpha}\rangle \hat G_{\alpha} \label{rho}\\
\hat H\left(t\right)=&\frac{\hbar}{2}\sum_{\alpha=0}^{15} \Gamma_{\alpha}\left(t\right)\hat G_{\alpha}\label{hamG}
\end{align}
\end{subequations}
It is crucial to note that Equation \eqref{rho} is universal in various contexts, establishing its fundamental nature. In contrast, Equation \eqref{hamG} pertains to a particular scenario relevant to our specific problem, utilizing traceless generators denoted as \(\hat{G}_{\alpha}\) along with the identity operator \(\hat{I}\), defined as $ |1\rangle \langle 1| + |2\rangle \langle 2| + |3\rangle \langle 3| + |4\rangle \langle 4|$, to describe the dynamics of the system in question. The generators \(\hat{G}_{\alpha}\) form a vector space with an inner product, defined by the trace relation \(\langle G_{\alpha} \rangle = \text{Tr}(\hat{\rho}(t) \hat{G}_{\alpha})\). This decomposition is done in terms of an orthonormal basis, where each generator corresponds to a specific direction in this vector space, and the coefficients \( \Gamma_{\alpha}(t) \) capture the time evolution of the system along these basis directions:
\begin{subequations}
\begin{align}
\langle G_{\alpha}\rangle=&Tr\left(\hat\rho\left(t\right)\hat G_{\alpha}\right) \label{avgG}\\
\hbar \Gamma_{\alpha}\left(t\right)=&Tr\left(\hat H\left(t\right)\hat G_{\alpha}\right)\label{coeh}
\end{align}
\end{subequations}
This formulation provides a clear view of the time evolution of the population of the system and the dynamics of coherence in the context of the vector space formed by the generators. The coefficient \(\langle G_{\alpha}\rangle\) represents the expectation value of the generator \(\hat{G}_{\alpha}\) within the given quantum state, providing insight into how this generator influences the dynamics of the system.  On the other hand, \(\Gamma_{\alpha}(t)\) typically characterizes the evolution of the system over time, capturing how the contributions of the generators change over time. It should be noted that Eq. \eqref{trace} indicates that the trace of the product of the generators is orthogonal, while Eq. \eqref{comuG} denotes the closure property.
Consequently, it follows that 
\begin{equation}
\begin{aligned}
\left[\hat H,\hat G_\alpha\right]=i \hbar\sum_\beta \hat G_\beta g_{\beta\alpha}, 
\label{HGcomm}
\end{aligned}
\end{equation}

Recall that the dynamics of an atomic system at the \(N\) level is represented by the density matrix \(\hat{\rho}\), which evolves in time according to the Liouville equation \cite{levine2011quantum, cohen1991quantum}. 
\begin{equation}
\begin{aligned}
\frac{\partial}{\partial t}\hat\rho=&\frac{i}{\hbar}\left[\hat\rho,\hat H\right]
\label{Liouville}
\end{aligned}
\end{equation}

By applying the Liouville equation and Equation \eqref{rho}, together with the closure relation, one can derive the equation of motion for the expectation value of an operator. This expectation value is denoted by the expression \(\langle G_{\alpha}\rangle = \mathrm{Tr}[\hat{\rho}(t) \hat{G}_{\alpha}]\) \cite{alhassid1977entropy}. The equation of motion for the generators of the \(\mathrm{SU(4)}\) Lie group can be readily obtained as follows:
\begin{equation}
\begin{aligned}
\frac{d}{dt}\langle G_{\alpha}\rangle=&-\sum_{\beta}\langle G_{\beta}\rangle  g_{\beta\alpha}
\label{vecG4AL}
\end{aligned}
\end{equation}
Where $\alpha, \beta = 1, 2, \dots, 15$, and $g_{\beta\alpha}$ denote (potentially complex) numerical coefficients. Equation \eqref{vecG4AL} comprises a set of coupled linear differential equations that fully describe the time evolution of the expectation values $\langle G_{\alpha}\rangle$, provided the initial values are known. 
Equation \eqref{vecG4AL} indicates that the rate of change of \( \langle G_\alpha \rangle \) is determined by a linear combination of the expectation values of other generators \( \langle G_\beta \rangle \), with the coefficients \( g_{\beta \alpha} \) encoding the coupling between them. To improve clarity, we explicitly relate \( g_{\beta \alpha} \) the parameters \( \Gamma_\alpha \) and the structure constants \( f_{\alpha \beta \gamma} \).  To this end, the system's Hamiltonian is expressed as a weighted sum of the generators:  
\begin{equation}
\begin{aligned}
\hat{H}(t) &= \frac{\hbar}{2} \sum_{\alpha=0}^{15} \Gamma_{\alpha}(t) \hat{G}_{\alpha},
\end{aligned}
\end{equation}
where \( \Gamma_{\alpha}(t) \) are time-dependent real parameters that determine the interaction strengths or external controls applied to the system. The time evolution of an operator in the Heisenberg picture is given by:  
\begin{equation}
\begin{aligned}
\frac{d}{dt} \hat{G}_\alpha &= \frac{i}{\hbar} [\hat{H}, \hat{G}_\alpha].
\end{aligned}
\end{equation}  
where \(\hat{H}\) represents the system's Hamiltonian, and \(\hat{G}_\alpha\) is a generator within the algebra governing the system. This equation encapsulates how the operator evolves with time under the influence of the system's Hamiltonian. To proceed with the evaluation of the time evolution, we substitute the explicit form of the commutator between the Hamiltonian and the operator \(\hat{G}_\alpha\). This commutator arises from equation~\eqref{HGcomm}:
\begin{equation}
\begin{aligned}
\relax[\hat{H}, \hat{G}_\alpha]
&= \relax \frac{\hbar}{2}
\sum_{\beta=0}^{15} \Gamma_{\beta}(t)
[\hat{G}_\beta, \hat{G}_\alpha].
\end{aligned}
\end{equation}
At this stage, we utilize the Lie algebra relation that the generators \(\hat{G}_\beta\) satisfy and are given in equation \eqref {comuG}, the commutator describes the interaction between different generators of the algebra.
Substituting this into the commutator, we obtain:  
\begin{equation}
\begin{aligned}
\relax[\hat{H}, \hat{G}_\alpha] &= i \hbar \sum_{\beta,\gamma=0}^{15} \Gamma_{\beta}(t) f_{\beta \alpha \gamma} \hat{G}_\gamma.
\end{aligned}
\end{equation}  
This expression highlights the coupling between different generators \(\hat{G}_\gamma\) due to the time-dependent coefficients \(\Gamma_{\beta}(t)\) and the structure constants \(f_{\beta \alpha \gamma}\), which encode the algebra's interactions.

To further analyze the dynamics, we now compare this result with the general form of the commutator between the Hamiltonian and the operators:
\begin{equation}
\begin{aligned}
\relax[\hat{H}, \hat{G}_\alpha] &= i \hbar \sum_{\beta=0}^{15} \hat{G}_\beta g_{\beta \alpha},
\end{aligned}
\end{equation}  
This step establishes how the coefficients \(g_{\beta \alpha}\) depend on the time-dependent coefficients \(\Gamma_{\gamma}(t)\) and the structure constants \(f_{\gamma \beta \alpha}\), thus linking the algebraic structure to the dynamics of the system, we identify the coefficients \( g_{\beta \alpha} \) as:  
\begin{equation}
\begin{aligned}
g_{\beta \alpha} &= \sum_{\gamma=0}^{15} \Gamma_{\gamma}(t) f_{\gamma \beta \alpha}.
\end{aligned}
\end{equation}  

Next, we focus on the dynamics of the expectation values of the operators. The time evolution of the expectation value \(\langle G_{\alpha} \rangle\) is given by:  
\begin{equation}
\begin{aligned}
\frac{d}{dt} \langle G_{\alpha} \rangle &= \frac{i}{\hbar} \langle [\hat{H}, \hat{G}_\alpha] \rangle.
\end{aligned}
\end{equation}  
This equation describes how the expectation value evolves in response to the commutator. Substituting the previously computed commutator, we obtain the following equation for the time evolution of the expectation values:
\begin{equation}
\begin{aligned}
\frac{d}{dt} \langle G_{\alpha} \rangle &= - \sum_{\beta=0}^{15} g_{\beta \alpha} \langle G_{\beta} \rangle.
\end{aligned}
\end{equation}  
Here, the coefficients \(g_{\beta \alpha}\) encapsulate the interaction between the different expectation values \(\langle G_{\beta} \rangle\), determining how the dynamics of one expectation value influences the others.

Finally, substituting the explicit expression for \(g_{\beta \alpha}\), we arrive at the complete evolution equation for the expectation values:  
\begin{equation}
\begin{aligned}
\frac{d}{dt} \langle G_{\alpha} \rangle &= - \sum_{\beta,\gamma=0}^{15} \Gamma_{\gamma}(t) f_{\gamma \beta \alpha} \langle G_{\beta} \rangle.
\end{aligned}
\end{equation}  
This equation provides a set of coupled differential equations that govern the time evolution of the expectation values of the generators \(\hat{G}_\alpha\). 
This formulation explicitly demonstrates how the dynamics of the system is governed by the interaction between external control parameters \( \Gamma_{\gamma}(t) \) and the intrinsic structure of the Lie algebra defined by the structure constants \( f_{\gamma \beta \alpha} \). The time evolution equation shows that the expectation values \( \langle G_{\alpha} \rangle \) evolve through a coupling mechanism dictated by these parameters, where \( \Gamma_{\gamma}(t) \) represents the externally imposed control functions, while the structure constants \( f_{\gamma \beta \alpha} \) encapsulate the inherent algebraic relationships between the generators. This result not only clarifies how the Hamiltonian influences the dynamics of the system but also reinforces the significance of the Lie algebra framework in shaping the temporal evolution of the expectation values. By explicitly incorporating the dependence on \( \Gamma_{\alpha} \) and \( f_{\alpha \beta \gamma} \), this formulation provides a comprehensive description of the behavior of the system, bridging the gap between external modulation and the fundamental mathematical structure that governs the generators. 

\section{Equation of motions}                                                                             
\label{sec:EQOM}
In the current context, generators are defined as linear combinations of operators $|j\rangle \langle k|$, as specified in the equations. \eqref{ujk}, \eqref{vjk}, and \eqref{wl}. Appendix \ref{su4} provides a comprehensive list of the complete set of 15 generators, along with the relevant structure constants $f_{ijk}$, which are detailed in Table \ref{su4c} \cite{Pfeifer2003,FTHIoeQME}.
\begin{subequations}
\begin{align}
\hat G_{m}=&|j\rangle\langle k|+|k\rangle\langle j|\label{ujk}\\
\hat G_{n}=&-i\left(|j\rangle\langle k|-|k\rangle\langle j|\right)\label{vjk}\\
\hat G_k=&-\sqrt{\frac{2}{l\left(l+1\right)}}\left(|j\rangle\langle j|-l|l+1\rangle\langle l+1|\right)\label{wl}
\end{align}
\label{gene}
\end{subequations}
Where $m = 1 \ldots 6$, $n = 7 \ldots 12$, and $k = 13 - 15$, with $i \neq j = 0 \ldots 3$ and $l = 0 \ldots 3$.

Following the procedures outlined in Section \ref{sec:SUND} diligently, one can derive the equations of motion for the expectation values of the generators. Organizing these equations into matrix form, we obtain the equation of motion for the coherence vector $\vec{G}$ of the four-level system, which is
\begin{equation}
\begin{aligned}
\frac{d}{dt}\vec G=\frac{1}{2}g_{15\times15}\vec G
\end{aligned}
\label{dvecG4}
\end{equation}
where $\vec G=(\langle G_1\rangle, \langle G_2\rangle,\cdots, \langle G_{15}\rangle)^T$, and 
\begin{equation}
\begin{aligned}
g_{15\times15}=
\begin{pmatrix}
0_{6\times6} & M_{6\times6} & 0_{6\times3}\\
-M_{6\times6}^T & 0_{6\times6} & K_{6\times3}\\
0_{3\times6} & -K_{3\times6}^T & 0_{3\times3}
\end{pmatrix}
\end{aligned}
\end{equation}
where \(0_{j \times k}\) with \(j, k = 3, 6\) represents a zero matrix of size \(j \times k\), and \(M^T\) and \(K^T\) indicate the transpose of the matrices \(M\) and \(K\), respectively.
\begin{widetext}
\begin{equation}
\begin{aligned}
M_{6\times6}=
\begin{pmatrix}
-\Delta  & 0 &  -\beta_1  & 0 & -\beta_2 & 0\\
0  & \Delta  &  \alpha_1  & 0  & \alpha_2 & 0 \\
-\beta_1  & \alpha_1 &  0  & -\alpha_2  & 0 & -\beta_2\\
0 & 0 & -\alpha_2 & -\Delta & -\alpha_1 & 0\\
\beta_2  & -\alpha_2 & 0 & \alpha_1  & 0 & \beta_2\\
0 & 0 & \beta_2 & 0 & \beta_2 & -\Delta 
\end{pmatrix},~&
K_{6\times3}= &
\begin{pmatrix}
-2\alpha_1  & 0 &  0 \\
\beta_1  & -\sqrt{3}\beta_1  & 0  \\
0  & 0 & 0 \\
-\beta_2  & -\frac{\sqrt{3}}{3}\beta_2  & -\frac{2\sqrt{6}}{3}\beta_2  \\
0 & 0 & 0\\
0  & -\frac{2\sqrt{3}}{3}\alpha_2  & \frac{2\sqrt{6}}{3}\alpha_2
\end{pmatrix}
\end{aligned}
\end{equation}
\end{widetext}

\begin{figure}[!htb]
\centering
(a)\includegraphics[width= 3.0 in]{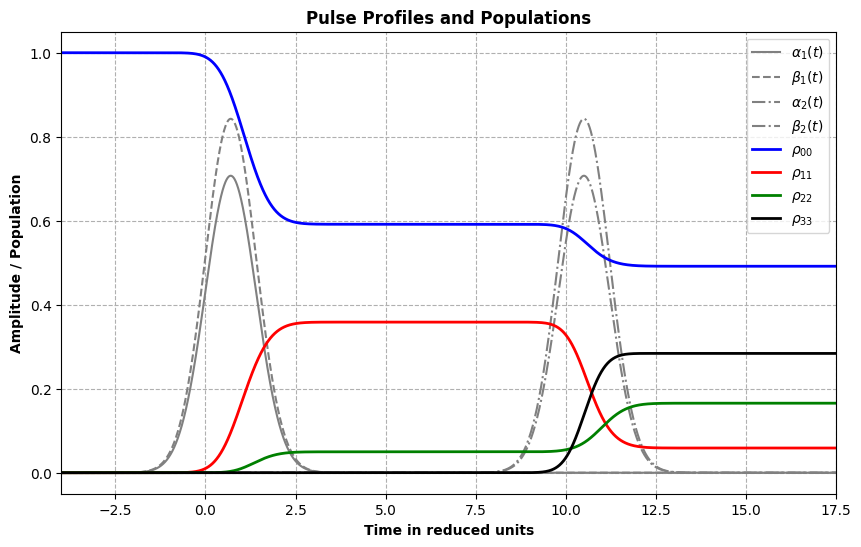}
(b)\includegraphics[width= 3.0 in]{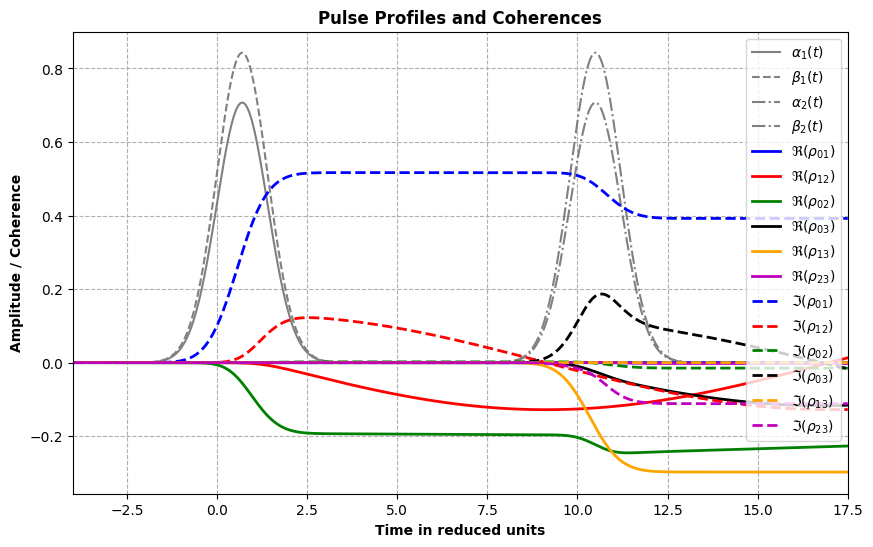}
\caption{Plots of the (a) Time evolution of the pulse profiles and the populations of the four states in a four-level diamond-shaped quantum system. The pulses drive population transitions between connected states. The population initially resides in a state $|0\rangle$, and as the pulses are applied, the population is cycled through the system, ultimately redistributing among the states. Time is presented in reduced units. (b) Time evolution of the pulse profiles and the real and imaginary parts of the coherences for a four-level diamond-shaped quantum system. The real ($\Re$) and imaginary ($\Im$) parts of the density matrix elements correspond to the coherences between the states, which evolve under the influence of the applied pulses. Coherence oscillations reflect population transfers and superpositions between these states. Time is shown in reduced units.}
\label{fig:Observables}
\end{figure}

The observable vector consists of 15 elements, each representing the expectation values of the generators. It is important to note that this observable vector serves to describe the states of the quantum system by providing average values of these generators. This method enables the tracking of the evolution of quantum states that can be observed experimentally \cite{parallelKsenia,DnstyQMDynamicsKsenia,dynamicsKsenia}.
The diamond-shaped four-level quantum system under consideration undergoes population transfers and coherence evolution in response to two pair time-dependent control pulses. The pulses drive transitions between states \(|0\rangle\), \(|1\rangle\), \(|2\rangle\), and \(|3\rangle\). The plot in Figure (\ref{fig:Observables}a) reveals the population dynamics, showing how the populations redistribute across the four levels due to external pulses. The plot in Figure (\ref{fig:Observables}b) shows the evolution of the coherences between these states, reflecting how the system enters superpositions and loses coherence as time progresses. 
\graphicspath{{Figures//}}
\FloatBarrier
Initially, the population is entirely in the state \(|0\rangle\), with \(\rho_{00} = 1.000\), and there is no population in the other states (\(|1\rangle\), \(|2\rangle\), or \(|3\rangle\)), with all coherence terms zero. This corresponds to the system starting fully localized in \(|0\rangle\), with no initial coherences between any of the states.
After applying pulses \(\alpha_1(t)\) and \(\beta_1(t)\), the populations redistribute. The probability amplitude \(\rho_{00}\) decreases to 0.592, while \(\rho_{11}\) increases to 0.359. This reflects a significant population transfer from state \(|0\rangle\) to state \(|1\rangle\), although state \(|0\rangle\) still holds the largest share of the population. The population in state \(|2\rangle\) becomes non-zero but remains small (\(\rho_{22} = 0.050\)), while state \(|3\rangle\) remains almost unoccupied (\(\rho_{33} = 0.000\)). 
After applying the next pair of pulses, that is, \(\alpha_2(t)\) and \(\beta_2(t)\), the population evolves further. Now, \(\rho_{00}\) decreases to 0.492, with \(\rho_{33}\) increasing to 0.284, indicating that the population of \(|0\rangle\) remains the largest, but that \(|3\rangle\) has become more populated. The population in \(|2\rangle\) also grows to 0.166, while \(|1\rangle\) decreases significantly to 0.059. This population redistribution reflects the influence of the stronger \(\beta_2(t)\), which favors the transitions between \(|3\rangle\) and \(|0\rangle\) over those between \(|2\rangle\) and \(|3\rangle\), as well as the smaller role of \(\alpha_2(t)\) in the transitions. 
\section{Dissipative Dynamics in a Four-Level System}
\label{sec:noise}
This section extends the previous analysis by incorporating dissipation, which accounts for relaxation effects arising from the system’s interaction with its environment. Previously, we derived the equations of motion under the assumption of an isolated system, neglecting environmental interactions. Here, we introduce the missing relaxation terms and formulate the evolution of the density matrix using a modified Liouville equation. The evolution of the elements of the density matrix $\rho_{mn} (t)$ is governed by the equation~\cite{mukamel1999principles,schirmer2004constraints,breuer2007theory}.
\begin{equation}
\begin{aligned}
\frac{d}{dt} \rho_{mn} (t) &= \frac{i}{\hbar} \left[\rho (t), H \right]_{mn} - \left(\Gamma \rho (t) \right)_{mn},
\end{aligned}
\end{equation}
where the first term describes unitary evolution due to the system’s Hamiltonian, while the second term represents dissipation. The unitary evolution is captured by the commutator
\begin{equation}
\begin{aligned}
\left[ \rho (t), H \right]_{mn} &= \sum_k \left( \rho_{mk} H_{kn} - H_{mk} \rho_{kn} \right),
\end{aligned}
\end{equation}
whereas the dissipation is expressed as
\begin{equation}
\left( \Gamma \rho (t) \right)_{mn} = \sum_{jk} \Gamma_{mn,jk} \rho_{jk}.
\end{equation}
Rewriting the equation of motion in the superoperator representation, we obtain
\begin{equation}
\begin{aligned}
\frac{d}{dt} \rho_{mn} = \frac{i}{\hbar} \sum_{jk} L_{mn,jk} \rho_{jk} - \sum_{jk} \Gamma_{mn,jk} \rho_{jk},
\end{aligned}
\end{equation}
where the first term,
\begin{equation}
L_{mn,jk} = H_{mj} \delta_{nk} - H^*_{nk} \delta_{mj},
\end{equation}
corresponds to the coherent evolution of the isolated system, while the second term represents the effects of the environment. Our goal in this section is to explicitly determine the structure of the dissipation superoperator \(\Gamma_{mn,jk}\), as the unitary evolution has already been established in previous discussions. The dissipation superoperator takes the form
\begin{equation}
\begin{aligned}
\Gamma_{mn,jk} &= \left[\frac{1}{2} \left( \Gamma_{mm,mm} + \Gamma_{nn,nn} \right) + \Gamma_{mn,mn} \right] \delta_{mj} \delta_{nk},
\end{aligned}
\end{equation}
where \(\Gamma_{mn,mn}\) represents the dephasing rate associated with the transition \(|m\rangle \leftrightarrow |n\rangle\), while \(\Gamma_{mm,mm}\) and \(\Gamma_{nn,nn}\) correspond to the decay rates of states \(|m\rangle\) and \(|n\rangle\), respectively. It is important to note that the total phase relaxation rate \(\Gamma_{mn}\) consists of two contributions: \(\Gamma_{mn}^p\), which accounts for decoherence due to population relaxation, and \(\Gamma_{mn}^d\), which represents pure dephasing without energy loss. The total decoherence rate is then given by
\begin{equation}
\begin{aligned}
\Gamma_{mn} &= \Gamma^p_{mn} + \Gamma^d_{mn},\\
\Gamma_{mn} &= \Gamma^d_{mn} + \frac{1}{2} \sum_{k \neq m,n} \left( \gamma_{km} + \gamma_{kn} \right), \quad m \neq n.
\end{aligned}
\end{equation}
Once these relaxation terms are determined, they can be directly incorporated into the framework of the system’s generators, providing a comprehensive description of the dissipative quantum dynamics.
\graphicspath{{Figures//}}
\begin{figure}[!htb]
\centering
(a)\includegraphics[width= 3.0 in]{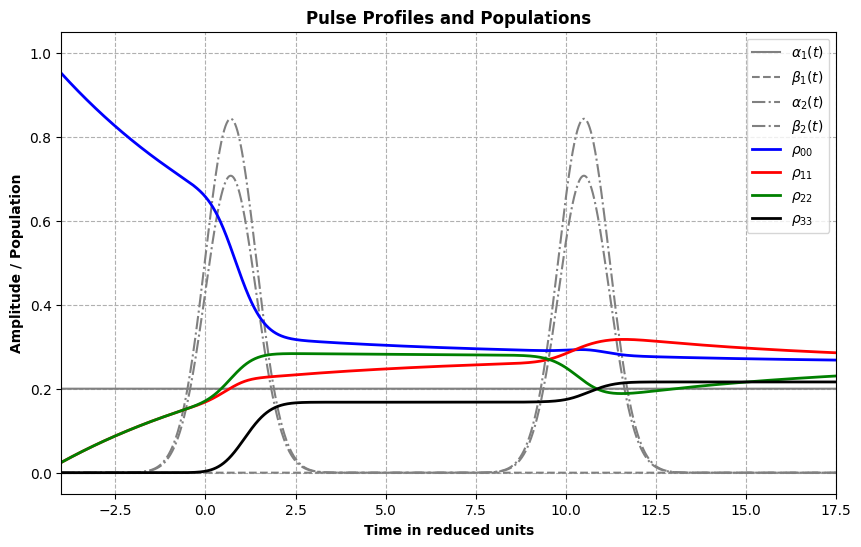}
(b)\includegraphics[width= 3.0 in]{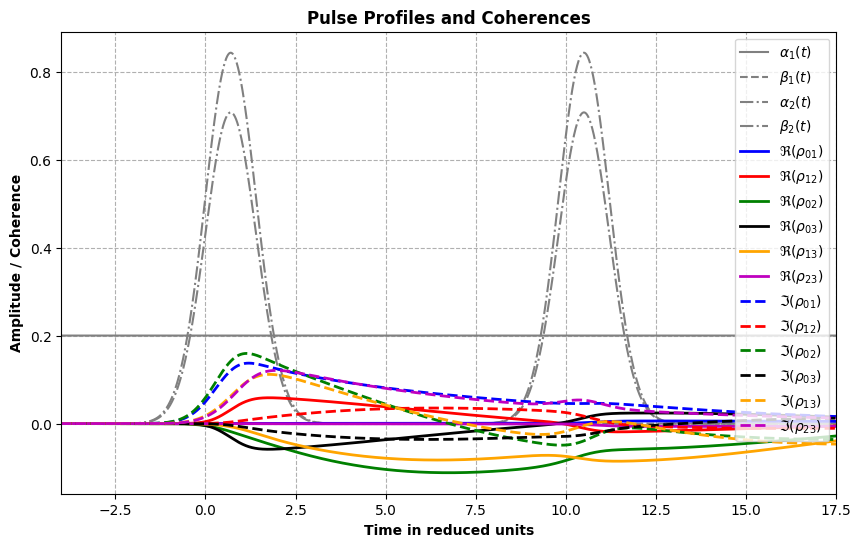}
\caption{(a) Time evolution of pulse profiles and state populations in a four-level diamond-shaped quantum system with noise. The applied pulses induce population transfers between connected states. Initially, the population is in state \( |0\rangle \), and as the pulses interact with the system, the population cycles through the various states, eventually reaching a new equilibrium distribution. Time is expressed in reduced units.  
(b) Time evolution of pulse profiles and the real and imaginary components of coherences in the same system. The real (\(\Re\)) and imaginary (\(\Im\)) parts of the density matrix elements describe coherences between states, which evolve due to the pulses. Oscillations in the coherences correspond to the transfer of population and the formation of superpositions between states. Time is displayed in reduced units.}
\label{fig:Observables_Noise}
\end{figure}
Figure (\ref{fig:Observables_Noise}) illustrates the impact of dissipation on population and coherence dynamics within a four-level quantum system. In the presence of noise, the evolution of the density matrix is influenced not only by the unitary dynamics governed by the Hamiltonian but also by the dissipation superoperator \(\Gamma\), which introduces relaxation and dephasing effects.  
Figure (\ref{fig:Observables_Noise}a) depicts the time evolution of the population terms \(\rho_{00}\), \(\rho_{11}\), \(\rho_{22}\), and \(\rho_{33}\) along with the external driving fields \(\alpha_1(t)\), \(\beta_1(t)\), \(\alpha_2(t)\), and \(\beta_2(t)\). Initially, the system is predominantly in the ground state, with \(\rho_{00} \approx 1\), indicating that the whole population resides in the state \(|0\rangle\). As the external pulses interact with the system, the population is transferred to the excited states. However, because of dissipation, coherent population oscillations do not persist indefinitely. Instead, relaxation processes drive the system toward equilibrium, leading to a gradual redistribution of the population. Over time, all population elements stabilize, highlighting how dissipation suppresses coherent transitions and leads to a steady-state configuration.  
Figure (\ref{fig:Observables_Noise}b) provides insight into the evolution of the coherence terms \(\rho_{mn}\), showing both their real and imaginary components as they respond to external driving and dissipation. At the onset, the external fields induce quantum coherence, consistent with unitary evolution. However, dissipation gradually suppresses these coherence terms, demonstrating the inevitable loss of quantum coherence due to environmental interactions. By the end of the simulation, coherence has largely decayed, leaving only minimal residual contributions.  
Together, these results reveal that in an open quantum system, observables diminish over time due to noise. In the absence of dissipation, coherent dynamics dominate, resulting in sustained population oscillations and persistent quantum coherence. However, once dissipation is introduced, the system undergoes decoherence and eventually reaches a steady-state population distribution resembling classical behavior. This underscores the profound impact of environmental interactions on quantum systems and highlights the necessity of mitigating decoherence in quantum computing and quantum information processing. For this reason, the quantum decision tree is constructed using population terms rather than coherence, since information stored in coherence is more susceptible to environmental disturbances.

In summary, for the isolated system that serves as the foundation for our tree proposal, the first pulse sequence, driven by \(\alpha_1\) and \(\beta_1\), primarily keeps the population in state \(|0\rangle\), with a smaller fraction transferred to state \(|1\rangle\) and an even smaller portion transferred to state \(|2\rangle\), while state \(|3\rangle\) remains nearly unoccupied. The coherence between states \(|0\rangle\) and \(|1\rangle\) is strong, characterized by a positive imaginary component, while the coherence between states \(|0\rangle\) and \(|2\rangle\) is weaker and dominated by a negative real component. Following the second pulse sequence, which involves \(\alpha_2\) and \(\beta_2\), the state \(|0\rangle\) continues to have the highest population, but the state \(|3\rangle\) surpasses the state \(|2\rangle\), while the state \(|1\rangle\) ends up with the smallest population.

To withstand loss of coherence, it is important to note that the implementation relies on systems such as rare earth ion-doped crystals, which have demonstrated exceptionally long decoherence times \cite{AlexanderRareEarth,Beil:2009aa,beil2011logic,heinze2010storage}. These materials are known for their ability to preserve quantum coherence over extended periods, making them ideal candidates for quantum information processing applications. The long decoherence times observed in these systems are attributed to their unique electronic structures and the ability to isolate the quantum states of the ions from environmental noise. As a result, these systems can maintain coherent superposition states for much longer than typical quantum systems, providing a robust foundation for the realization of quantum decision trees and other quantum computational models.

In conclusion, the system exhibits a dynamic evolution in both coherence and population, driven by time-dependent pulses with synchronized temporal profiles but differing amplitudes. The interplay between these pulses dictates the extent of population transfer and the degree of coherence between the connected states, leading to the observed patterns in the results.
\section{Towards Quantum Decision Tree Model}                                                                             
\label{sec:decsiontree}
In this work, we introduce a framework that facilitates the way towards the implementation of a quantum decision tree model through a pump-probe scheme, enhancing classical decision-making frameworks by incorporating quantum mechanical principles. The proposed tree shows a distinct computational advantage over traditional decision trees, particularly in high-dimensional decision-making tasks where classical methods face significant complexity challenges \cite{parallelKsenia, remaclepar, Grove, QMDT98}. 

To reiterate, in our approach, the system is initialized in a pure quantum state, denoted \(|0
\rangle\), serving as the root node of the decision tree. This root state has a probability amplitude of 1.0, ensuring absolute certainty in the starting configuration. As the system evolves through controlled interactions, such as electromagnetic pulses, it transitions into new quantum states. The decision nodes correspond to these quantum states, and the edges between them represent the transition probabilities dictated by the control parameters \(\alpha_j(t)\) and \(\beta_j(t)\) \cite{parallelKsenia, remaclepar}.

However, we focus on the interaction that allows the initial state \(|0
\rangle\) to evolve into four possible states: \(|0\rangle, |1\rangle, |2\rangle, |3\rangle\). It must be pointed out that, in addition to these transitions, coherence terms exist between these states, significantly increasing the complexity of the decision tree. Unlike classical decision trees, where transitions are governed by simple branching probabilities, quantum interference modulates these probabilities based on the relative phases of quantum states.

Mathematically, the probability of transitioning from state \(|j\rangle\) to state \(|k\rangle\) can be computed using the unitary evolution operator \(U\):
\begin{equation}
    \begin{aligned}
        P_{jk} = \left|\sum_{m} U_{jm} U_{mk}^{*} \right|^2.
    \end{aligned}
\end{equation}
This expression highlights the role of interference, since probability amplitudes are added coherently rather than classically. 
Beyond population dynamics, coherence effects significantly impact the evolution of the system. Coherence, represented by the off-diagonal elements of the density matrix \(\rho_{jk}\), determines the quantum correlations between states. Unlike classical decision trees that follow a deterministic trajectory, our quantum model simultaneously explores multiple decision paths due to superposition. The presence of coherence terms leads to interference effects that influence the probabilities of transition. For example, the probability of transitioning from \(|0\rangle\) back to itself after two interactions is given by:
\begin{equation}
    \begin{aligned}
P_{00}^{\text{total}} = \sum_{m,n} U_{0m} U_{mn} U_{n0}^{*}.
\end{aligned}
\end{equation}
The sum of all intermediate states accounts for all possible pathways, including those that constructively or destructively interfere. Coherence terms indicate that transitions between states are not independent, but rather entangled through quantum correlations, further complicating the decision process. This is observed in our numerical simulations, where population dynamics and coherence effects exhibit strong temporal variations under pulsed excitations, as shown in Figures (\ref{fig:Observables}) and (\ref{fig:Observables_Noise}).

To illustrate this concept more concretely, consider the transition probability of reaching the state \(|0\rangle\) after two interactions. This probability is given by the sum of the contributions from all the paths leading to \(|0\rangle\). These paths include sequences like \(0 \rightarrow 0 \rightarrow 0\), \(0 \rightarrow 1 \rightarrow 0\), \(0 \rightarrow 2 \rightarrow 0\), and \(0 \rightarrow 3 \rightarrow 0\), with each contribution determined by the product of the corresponding transition amplitudes. Mathematically, the total probability is expressed as:
\begin{equation}
    \begin{aligned}
P_{00}^{\text{total}} = (0.59 \times 0.49) + (0.36 \times 0.49) + (0.05 \times 0.49) + (0.00 \times 0.49) = 0.49.
\end{aligned}
\end{equation}

By adding these contributions, we obtain the probability of transitioning to the state \(|0\rangle\), which accounts for all possible paths leading to this state.

\subsection{Computational Advantages and Quantum Speed up} 

The key advantage of quantum decision trees is their ability to exploit quantum parallelism, reducing the computational resources required for exploring decision paths. This parallelism is a result of the principle of quantum superposition, which allows the system to be in multiple states simultaneously. Although classical decision trees evaluate decision paths one at a time, our quantum model processes all paths simultaneously, speeding up the decision-making process \cite{Grove}. This behavior mirrors well-established quantum search algorithms, like Grover's algorithm, which provides quadratic speed up for database searches. In our model, the probability distribution adheres to similar principles of quantum search, with amplitude amplification enhancing the likelihood of selecting the optimal decision pathways. The search efficiency is defined by:
\begin{equation}
    \begin{aligned}
T = O\left( \sqrt{N} \right),
\end{aligned}
\end{equation}
where \(N\) is the number of possible decision states. This result confirms that quantum decision trees provide a fundamental computational advantage over their classical counterparts \cite{QMDT98}.

\subsection{Scalability and Practical Implementation} 

Extending this framework to larger quantum systems demonstrates scalability, as the number of quantum states and decision nodes grows exponentially. The general state of an \(N\)-level quantum system can be written as $|\psi\rangle = \sum_{j=0}^{N-1} c_j |j\rangle,$ where \(c_j\) are complex probability amplitudes that evolve under unitary transformations. This expansion enables efficient exploration of complex decision landscapes, making quantum decision trees well-suited for large-scale optimization and machine learning applications. These transition probabilities are not uniformly distributed, as they are influenced by the specific interaction parameters and quantum interference effects. Quantum mechanics dictates that after interaction, the system remains in a superposition of possible states. The new state after the first pair of pulses can be expressed as \cite{remacle2006all}:
\begin{equation}
|\Psi'\rangle = c_0 |0\rangle + c_1 |1\rangle + c_2 |2\rangle + c_3 |3\rangle
\end{equation}
where \( c_i \) are complex probability amplitudes satisfying the normalization condition:
\begin{equation}
\sum_i |c_i|^2 = 1.
\end{equation}
The probability of measuring the system in a particular state \(|i\rangle\) is given by \( P_i = |c_i|^2 \), which corresponds to the color-coded transition probabilities in the quantum tree. After the first pulse sequence, the system undergoes a second unitary transformation:
\begin{equation}
|\Psi''\rangle = \hat{U} |\Psi'\rangle.
\end{equation}
Expanding this expression yields:
\begin{equation}
|\Psi''\rangle = \sum_{i,j} c_{ij} |ij\rangle,
\end{equation}
where \( c_{ij} \) represents the new probability amplitudes after the second interaction, and the states \(|ij\rangle\) correspond to the leaf nodes in the quantum decision tree. The successive application of unitary transformations is fundamental to quantum computing and quantum decision-making, where controlled interactions dictate the final measurement outcome. Notably, although not explicitly represented as transition nodes in the quantum tree, coherence propagation occurs after each pulse interaction, influencing the system’s dynamics.
However, practical implementation requires addressing decoherence and noise effects in quantum systems. The robustness of the model can be enhanced using error-correcting codes and decoherence-free subspaces, ensuring reliable decision making even in noisy quantum environments \cite{DynamicalSupression, PRXQuantumNoise, PShorDecoh,SteaneErrorQM}.

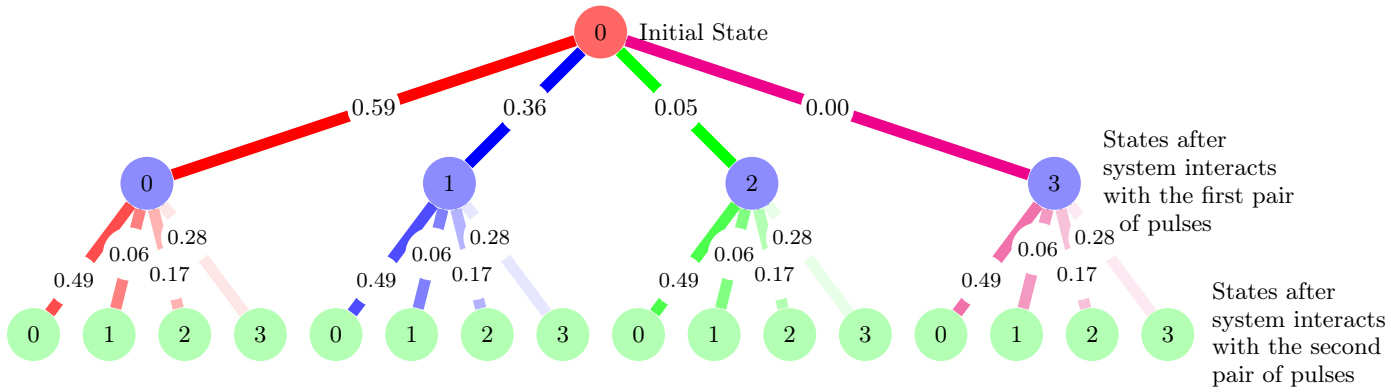
\begin{figure}
\centering    
\begin{tikzpicture}
  [scale=1.0, 
   level distance=20mm, 
   every node/.style={circle, minimum size=7mm, inner sep=1pt, font=\small}, 
   level 1/.style={sibling distance=40mm, nodes={fill=blue!45}}, 
   level 2/.style={sibling distance=10mm, nodes={fill=green!30}}, 
   edge from parent/.style={draw, -latex, thick}]
  
  \node[fill=red!60] (root) {0} 
    child {node (A0) {0}
      child {node (A0B0) {0}}
      child {node (A0B1) {1}}
      child {node (A0B2) {2}}
      child {node (A0B3) {3}}
    }
    child {node (A1) {1}
      child {node (A1B0) {0}}
      child {node (A1B1) {1}}
      child {node (A1B2) {2}}
      child {node (A1B3) {3}}
    }
    child {node (A2) {2}
      child {node (A2B0) {0}}
      child {node (A2B1) {1}}
      child {node (A2B2) {2}}
      child {node (A2B3) {3}}
    }
    child {node (A3) {3}
      child {node (A3B0) {0}}
      child {node (A3B1) {1}}
      child {node (A3B2) {2}}
      child {node (A3B3) {3}}
    };

  \path[draw=red, line width=1.5mm] (root) -- (A0) node[midway, fill=white] {0.59};
  \path[draw=blue, line width=1.5mm] (root) -- (A1) node[midway, fill=white] {0.36};
  \path[draw=green, line width=1.5mm] (root) -- (A2) node[midway, fill=white] {0.05};
  \path[draw=magenta, line width=1.5mm] (root) -- (A3) node[midway, fill=white] {0.00};

  \path[draw=red!70, line width=1.5mm] (A0) -- (A0B0) node[pos=0.7, fill=white, font=\footnotesize] {0.49};
  \path[draw=red!50, line width=1.5mm] (A0) -- (A0B1) node[pos=0.45, fill=white, font=\footnotesize] {0.06};
  \path[draw=red!30, line width=1.5mm] (A0) -- (A0B2) node[pos=0.65, fill=white, font=\footnotesize] {0.17};
  \path[draw=red!10, line width=1.5mm] (A0) -- (A0B3) node[pos=0.3, fill=white, font=\footnotesize] {0.28};

  \path[draw=blue!70, line width=1.5mm] (A1) -- (A1B0) node[pos=0.7, fill=white, font=\footnotesize] {0.49};
  \path[draw=blue!50, line width=1.5mm] (A1) -- (A1B1) node[pos=0.45, fill=white, font=\footnotesize] {0.06};
  \path[draw=blue!30, line width=1.5mm] (A1) -- (A1B2) node[pos=0.65, fill=white, font=\footnotesize] {0.17};
  \path[draw=blue!10, line width=1.5mm] (A1) -- (A1B3) node[pos=0.3, fill=white, font=\footnotesize] {0.28};

  \path[draw=green!70, line width=1.5mm] (A2) -- (A2B0) node[pos=0.7, fill=white, font=\footnotesize] {0.49};
  \path[draw=green!50, line width=1.5mm] (A2) -- (A2B1) node[pos=0.45, fill=white, font=\footnotesize] {0.06};
  \path[draw=green!30, line width=1.5mm] (A2) -- (A2B2) node[pos=0.65, fill=white, font=\footnotesize] {0.17};
  \path[draw=green!10, line width=1.5mm] (A2) -- (A2B3) node[pos=0.3, fill=white, font=\footnotesize] {0.28};

  \path[draw=magenta!70, line width=1.5mm] (A3) -- (A3B0) node[pos=0.7, fill=white, font=\footnotesize] {0.49};
  \path[draw=magenta!50, line width=1.5mm] (A3) -- (A3B1) node[pos=0.4, fill=white, font=\footnotesize] {0.06};
  \path[draw=magenta!30, line width=1.5mm] (A3) -- (A3B2) node[pos=0.65, fill=white, font=\footnotesize] {0.17};
  \path[draw=magenta!10, line width=1.5mm] (A3) -- (A3B3) node[pos=0.3, fill=white, font=\footnotesize] {0.28};
  
  \node at (root.east) [right=1mm] {Initial State};
  \node at (A3.east) [right=1.2mm, text width=4cm, align=left] {States after \\ system interacts \\ with the first pair\\~ of pulses};
  \node at (A3B3.east) [right=0.6mm, text width=4cm, align=left] {States after \\ system interacts \\ with the second\\ pair of pulses};
\end{tikzpicture}
\caption{The proposed quantum tree with colored paths}
\end{figure}
\section{Conclusion}
\label{sec:conclusion}
In conclusion, we have examined a four-level diamond-shaped quantum system where interlevel interactions are facilitated by Stokes and pump pulses. Specifically, states $ |0\rangle, |3\rangle $ are connected by Stokes lasers $\beta_2(t)$, while states $ |1\rangle, |2\rangle $ are coupled by $\beta_1(t)$ respectively. Pump lasers $\alpha_1(t)$ couple states $ |0\rangle, |1\rangle $, and $\alpha_2(t)$ couples $ |2\rangle, |3\rangle $, respectively. By assuming that the pulses exhibit identical time dependencies while possibly differing in amplitude, we were able to effectively redistribute the population that initially resided in the state \( |0\rangle \) across all available states within the system. 

This approach allows us to manipulate the population of our four-level system in such a manner that it engages in a dynamic process reminiscent of quantum decision trees. In this context, the system evolves based on a series of quantum operations that determine the probabilities of transitioning between states, ultimately leading to distinct outcomes depending on the amplitudes and phases of the applied pulses. The behavior of this setup illustrates how quantum mechanics can be harnessed to perform complex decision-making tasks through coherent superpositions and interference effects among the different states.

In essence, the quantum decision tree model takes advantage of the inherent parallelism and interconnectivity of quantum systems, offering a computational edge over classical decision trees, especially in scenarios that involve large datasets or complex decision paths. This capability underpins many quantum algorithms that aim to outperform their classical counterparts in tasks such as optimization, search, and machine learning.

The quantum tree serves as a model for quantum decision-making, where each branch represents a probabilistic outcome conditioned on prior interactions. This structure parallels decision-making in classical machine learning but harnesses quantum superposition and interference, enabling exponentially richer state representations.

\appendix
\renewcommand{\thesection}{A.\arabic{section}}
\numberwithin{equation}{section}
\section{Generators of $\mathrm{SU(4)}$}
\label{su4}
The $\mathrm{SU(4)}$ generators utilized in our analysis are: \cite{Pfeifer2003}
\begin{widetext}
\begin{equation}
\begin{aligned}
\hat G_1=&\begin{pmatrix}
0 & 1 & 0 & 0\\
1 & 0 & 0 & 0\\
0 & 0 & 0 & 0\\
0 & 0 & 0 & 0\end{pmatrix},     &    \hat G_2=&\begin{pmatrix}
0 & 0 & 0 & 0\\
0 & 0 & 1 & 0\\
0 & 1 & 0 & 0\\
0 & 0 & 0 & 0\end{pmatrix},       & \hat G_3=&\begin{pmatrix}
0 & 0 & 1 & 0\\
0 & 0 & 0 & 0\\
1 & 0 & 0 & 0\\
0 & 0 & 0 & 0\end{pmatrix},       \\
\end{aligned}
\end{equation}
\begin{equation}
\begin{aligned}
\hat G_4=&\begin{pmatrix}
0 & 0 & 0 & 1\\
0 & 0 & 0 & 0\\
0 & 0 & 0 & 0\\
1 & 0 & 0 & 0\end{pmatrix},       & \hat G_5=&\begin{pmatrix}
0 & 0 & 0 & 0\\
0 & 0 & 0 & 1\\
0 & 0 & 0 & 0\\
0 & 1 & 0 & 0\end{pmatrix},       & \hat G_6=&\begin{pmatrix}
0 & 0 & 0 & 0\\
0 & 0 & 0 & 0\\
0 & 0 & 0 & 1\\
0 & 0 & 1 & 0\end{pmatrix},       \\ 
\hat G_7=&\begin{pmatrix}
0 & i & 0 & 0\\
-i & 0 & 0 & 0\\
0 & 0 & 0 & 0\\
0 & 0 & 0 & 0\end{pmatrix},     &    \hat G_8=&\begin{pmatrix}
0 & 0 & 0 & 0\\
0 & 0 & i & 0\\
0 & -i & 0 & 0\\
0 & 0 & 0 & 0\end{pmatrix},       & \hat G_9=&\begin{pmatrix}
0 & 0 & i & 0\\
0 & 0 & 0 & 0\\
-i & 0 & 0 & 0\\
0 & 0 & 0 & 0\end{pmatrix},       \\
\end{aligned}
\end{equation}
\begin{equation}
\begin{aligned}
\hat G_{10}=&\begin{pmatrix}
0 & 0 & 0 & i\\
0 & 0 & 0 & 0\\
0 & 0 & 0 & 0\\
-i & 0 & 0 & 0\end{pmatrix},       &  \hat G_{11}=&\begin{pmatrix}
0 & 0 & 0 & 0\\
0 & 0 & 0 & i\\
0 & 0 & 0 & 0\\
0 & -i & 0 & 0\end{pmatrix},       &  \hat G_{12}=&\begin{pmatrix}
0 & 0 & 0 & 0\\
0 & 0 & 0 & 0\\
0 & 0 & 0 & i\\
0 & 0 & -i & 0\end{pmatrix},  \\
\hat G_{13}=&\begin{pmatrix}
-1 & 0 & 0 & 0\\
0 & 1 & 0 & 0\\
0 & 0 & 0 & 0\\
0 & 0 & 0 & 0\end{pmatrix},     &    \hat G_{14}=&\frac{1}{\sqrt{3}}\begin{pmatrix}
-1 & 0 & 0 & 0\\
0 & -1 & 0 & 0\\
0 & 0 & 2 & 0\\
0 & 0 & 0 & 0\end{pmatrix},       &  \hat G_{15}=&\frac{1}{\sqrt{6}}\begin{pmatrix}
-1 & 0 & 0 & 0\\
0 & -1 & 0 & 0\\
0 & 0 & -1 & 0\\
0 & 0 & 0 & 3\end{pmatrix},        
 \end{aligned}
\end{equation}
\end{widetext}

with their  structure constants is given by 
\begin{table}[htbp]
\begin{center}
\begin{tabular}{r|r|r|r|r|r|r|r}
\hline
ijk & 1,2,9  & 1,3,8 & 1,4,11 & 1,5,10 & 1,7,13 & 2,3,7 & 2,5,12  \\
\hline
$f_{ijk}$ & $-\frac{1}{2}$ & $-\frac{1}{2}$ & $-\frac{1}{2}$ &  $-\frac{1}{2}$ & 1 & $\frac{1}{2}$ & $-\frac{1}{2}$\\
\hline
\hline
ijk & 2,6,11  & 2,8,13 & 2,8,14 & 3,4,12 & 3,6,10 & ,3,9 13 & 3,9,14  \\
\hline
$f_{ijk}$ & $-\frac{1}{2}$ & $-\frac{1}{2}$ & $\frac{\sqrt{3}}{2}$ &  $-\frac{1}{2}$ & $-\frac{1}{2} $& $\frac{1}{2}$ & $\frac{\sqrt{3}}{2}$\\
\hline
\hline
ijk & 4,5,7  & 4,6,9 & 4,10,13 & 4,10,14 & 4,10,15 & 5,6,8 & 5,11,13  \\
\hline
$f_{ijk}$ & $-\frac{1}{2}$ & $-\frac{1}{2}$ & $\frac{1}{2}$ &  $\frac{1}{2\sqrt{3}}$ & $\frac{\sqrt{2}}{\sqrt{3}} $& $-\frac{1}{2}$ & $-\frac{1}{2}$\\
\hline
\hline
ijk & 5,11,14  & 5,11,15 & 6,12,14 & 6,12,15 & 7,8,9 & 7,10,11 & 8,11,12  \\
\hline
$f_{ijk}$ & $\frac{1}{2\sqrt{3}}$ & $\frac{\sqrt{2}}{\sqrt{3}} $ & $-\frac{1}{\sqrt{3}}$ &  $\frac{\sqrt{2}}{\sqrt{3}} $ & $\frac{1}{2} $& $-\frac{1}{2}$ & $-\frac{1}{2}$ \\
\hline
\hline
ijk & 9,10,12    \\
\hline
$f_{ijk}$ & $-\frac{1}{2}$ \\
\hline
\end{tabular}
\caption{\label{tab:tc}value of the surviving structure constants}
\label{su4c}
\end{center}
\end{table}

\bibliography{Implementation_of_QMDT/Implementation_of_QMDT1}

\end{document}